# Exploring the behavior of vanadium under high-pressure and high-temperature conditions

D. Errandonea[1], S. G. MacLeod[2,3], L. Burakovsky[4], D. Santamaria-Perez[1], J. E. Proctor[3,†], H. Cynn[5], and M. Mezouar[6]

[1]*Departamento de Física Aplicada-ICMUV, Universidad de Valencia, MALTA Consolider Team, Edificio de Investigación, C/Dr. Moliner 50, 46100 Burjassot, Valencia, Spain*

[2]*Atomic Weapons Establishment, Aldermaston, Reading, RG7 4PR, United Kingdom*

[3]*SUPA, School of Physics and Astronomy, and Centre for Science at Extreme Conditions, The University of Edinburgh, Edinburgh, EH9 3FD, United Kingdom*

[4]*Theoretical Division, Los Alamos National Laboratory, Los Alamos, New Mexico 87545, USA*

[5]*Physics Division, Lawrence Livermore National Laboratory, Livermore, CA 94550, USA*

[6]*ID27 Beamline, European Synchrotron Radiation Facility, 71 Avenue des Martyrs, 38000 Grenoble, France*

[†]*Present address: School of Computing, Science, and Engineering, University of Salford, Manchester M5 4WT, UK*

**Abstract:** We report a combined experimental and theoretical study of the melting curve and the structural behavior of vanadium under extreme pressure and temperature. We performed powder x-ray diffraction experiments up to 120 GPa and 4000 K, determining the phase boundary of the bcc-to-rhombohedral transition and melting temperatures at different pressures. Melting temperatures have also been established from the observation of temperature plateaus during laser heating, and the results from the density-functional theory calculations. Results obtained from our experiments and calculations are fully consistent and lead to an accurate determination of the melting curve of vanadium. These results are discussed in comparison with previous studies. The melting temperatures determined in this study are higher than those previously obtained using the speckle method, but also considerably lower than those obtained from shock-wave experiments and linear muffin-tin orbital calculations. Finally, a high-pressure high-temperature equation of state up to 120 GPa and 2800 K has also been determined.

1. Introduction

Since the early study by the Mainz group reported nearly two decades ago [1], the melting of transition metals at high pressure (P) has been the focus of several research groups. The main motivation for these studies was to resolve the discrepancies in the values of melting temperatures (T) from diamond anvil cell (DAC) and shock-wave (SW) experiments, and density-functional theory (DFT) calculations. Most of the studies have focused on iron (Fe) [2-4], tantalum (Ta) [5-7], and molybdenum (Mo) [8-10]. Several hypotheses have been proposed to explain the apparent disagreements [11-14]. In the case of Mo, a recent study [9] has shown that microstructure formation could be the cause of the underestimation of melting temperature in the early studies.

Unlike other transition metals, vanadium (V) has scarcely been studied at high pressure (HP) and high temperature (HT). In fact, its melting T has only been calculated at one pressure point above ambient P [15]. This calculated melting T is higher than those estimated from both DAC [1] and SW [16] experiments, which are themselves separated by more than 1000 K at a pressure of 100 GPa. Clearly, further efforts should be dedicated to the study of the behavior of V under HP-HT conditions. The accurate determination of the melting curve of V is not only important from a fundamental physics point of view, but also from a technological one, since most of the V produced is for usage in metallurgy as a steel additive, to increase the strength of steel.

Furthermore, V is the only body-centered cubic (bcc)-structured transition metal for which a phase transition has been reported below 100 GPa [17]. The transition from bcc to a rhombohedral phase at 69 GPa appears to be strongly affected by non-hydrostatic conditions [17, 18]. According to DFT calculations, at HT the bcc-rhombohedral transition is predicted to be reversible [19]. However, this prediction has yet to be tested by experiments. This, and the findings described above indicate the appropriateness of performing HP-HT studies on V.

In this work, we report synchrotron powder x-ray diffraction (XRD) studies on V up to 120 GPa and 4000 K. These studies are combined with experiments where the observation of temperature plateaus is used as a melting diagnostic. DFT calculations are carried out to validate the experimental results. These studies have allowed us to accurately determine the melting curve and the bcc-rhombohedral phase boundary of V, and its HP-HT equation of state (EOS). The reported results will be discussed and compared with previous DAC and SW experiments.

## 2. Methods

Angle-dispersive XRD measurements were performed at beamline ID27 of the European Synchrotron Radiation Facility (ESRF) in five different samples to reduce the chances of chemical reactions. Platelets of V (99.99% purity, Aldrich) of around 25 μm in diameter and 5 μm thick were loaded into DACs with anvil culets of 200-280 μm. We used rhenium gaskets pre-indented to a 30 μm thickness. NaCl was the pressure-transmitting medium and thermal insulator, except for the experiment carried out at 120 GPa, for which the pressure medium was MgO. Pressure was determined from the HP-HT EOS of NaCl [20] or MgO [20] and tungsten (W) [21], which was loaded together with V. Agreement between different pressure scales was better than 2 GPa and the pressures given in the manuscript are the average of the pressures obtained from different pressure standards.

Pressure was increased at room temperature (RT) up to the target pressure, and then the samples were gradually heated under constant load. For the experiments we used a double-side laser heating system equipped with infrared YAG lasers, previously described [22]. The temperature was determined from the thermal emission of the samples [23]. The x-ray beam ($\lambda$ = 0.3738 Å) was focused on a 2 x 2 $\mu m^2$ area. During each run, XRD and thermal emission spectra were recorded simultaneously every 2 seconds. XRD was measured using a MAR CCD detector. The indexing and refinement of the powder XRD patterns were performed using PowderCell [24].

In addition to synchrotron experiments, a second set of experiments was performed to determine melting based upon the appearance of a plateau in the temperature as a function of applied laser power [2, 6]. In this case, single-side laser heating was used [25], NaCl was the pressure medium, temperatures were measured using the same technique employed in HP-HT synchrotron experiments, and pressures were measured using the ruby fluorescence technique at RT [26] and corrected by taking account of the thermal pressure [27].

We have also calculated the melting curve of V using the Z method [28] implemented with quantum molecular dynamics (QMD) simulations using VASP. These simulations are based on DFT. They were performed using the Perdew-Burke-Ernzerhof (PBE) exchange-correlation functionals and considering thirteen valence electrons per atom (3s, 3p, 3d, and 4s orbitals). We used a 432-atoms (6 × 6 × 6) supercell with a single Γ-point. For such a large supercell, full energy convergence (to ≤ 1 meV/atom) is already achieved, which was tested by running simulations with denser k-point meshes (2 × 2 × 2 and 3 × 3 × 3) and comparing their result with that of the run with a single Γ-point. We carried out the calculation of six melting points corresponding to six different unit cells with lattice constants of (in Å) 3.35, 3.15, 2.90, 2.75, 2.60, and 2.45. The duration of the runs was 15000-25000 time-steps of 1 fs each. The P-T coordinates of the six melting points that these QMD simulations produced are (P in GPa, T in K) (-14.7, 1540), (6.7, 2410), (40.1, 3060), (101, 4060), (207, 5230), (390, 6870). The error in T is about half of the initial T increment for a sequence of runs at a given fixed supercell volume. This increment was 250 K for the two lowest-P points, and 312.5 K for the remaining four points; hence the corresponding errors are (approximately) ±125 K and ±160 K, respectively. The error in P is negligibly small: less than 1 GPa for the two lowest-P points, and a few GPa for the remaining four points.

## 3. Results and discussion

Before presenting the results, we would like to state that in our experiments we did not find any clear evidence of chemical reactions with the pressure media [29] or with carbon migrating from the diamond anvils [30]. Fig. 1 shows a selection of diffraction patterns measured at different P-T conditions. At RT and 29 GPa the pattern can be unambiguously identified as the bcc structure of V, with a unit-cell parameter of 2.898(1) Å, plus the B1 structure pattern from NaCl and bcc pattern for W. The small residuals obtained after a Rietveld refinement (shown in the figure) support the identification of bcc-V. Under compression, at RT the bcc phase of V is observed up to 53 GPa. At higher pressures, the (110) and (211) peaks of V considerably broaden, especially the (211) peak. However, the width of the (200) is unaffected. This is illustrated in Fig. 1 by the pattern measured at 64 GPa (notice that in this pattern NaCl has transformed from the B1 to the B2 phase). The broadening of the two peaks of V can be accounted for by assuming a rhombohedral distortion of bcc as proposed by Ding *et al.* [17]. As shown by the Rietveld refinement in Fig. 1, the pattern measured at RT and 64 GPa can be assigned to a rhombohedral structure belonging to space group R-3m with unit-cell parameters *a* = 2.431(1) Å and $\alpha$ = 109.47(5)º. The rhombohedral phase is observed in our measurements at RT up to 120 GPa. At HT, the rhombohedral structure is observed up to 1560 K at 64 GPa (see Fig. 1) and up to 1700 K at 120 GPa. At higher temperatures the bcc phase is recovered as can be seen by the pattern measured at 64 GPa and 1840 K (notice that the three peaks of V are sharp again). This observation is consistent with both the predictions made by Wang *et al.* [18] and the calculations of Landa *et al.* [15] who find that bcc-V is stable at 75 GPa and 2500 K. This suggests that the phonon anomalies triggering the rhombohedral lattice distortion in V at HP [31] can be canceled at HT by effects from phonon-phonon scattering [32].

In Fig. 2 we present a selection of XRD patterns measured in a heating run at 32 GPa. Up to 2560 K, all the peaks of bcc V, W, and B2 NaCl are present. Beyond this temperature, but below 2930 K, we observed two phenomena: 1) a rapid recrystallization of V, with the peaks associated with it showing preferred orientations, which randomly change from one temperature to the next; and 2) the melting of NaCl, which causes the near disappearance of NaCl peaks (they do not completely disappear because NaCl in contact with diamonds is cooler than NaCl in contact with the sample, thus only a fraction of it is molten). In particular, at 2790 K NaCl is molten, which is in good agreement with the melting curve reported by Boehler *et al.* [33]. At this temperature, V is continuously recrystallizing. We also noticed that the V peaks become broader, a phenomenon that has also been observed in Mo below the melting temperature [9], and which is probably related to pre-melting effects. At 2930 K, we observed the complete disappearance of the peaks of V and the increase of the background. In addition, we then observed that beyond this point, the estimated temperature of V did not increase with further increases in laser power of up to 20%. These observations can be interpreted as the onset of melting [34-36]. In Fig. 3, we zoom in and focus on the XRD patterns in the 8º - 12º region to ease the identification of the background increase.

Using the method described above we have been able to determine the melting temperature of V at 27, 32, 53, and 64 GPa. At 120 GPa, V remained solid up to the highest temperature reached at this pressure (3750 K), but recrystallization of V was observed from 3300 to 3750 K. We have also determined the melting temperature from experiments where melting was detected using the temperature plateau criterion [34]. These melting temperatures were obtained at 40, 58, and 85 GPa. The results are summarized in Fig. 4, where we present a P-T phase diagram for V. We not only include the melting curve from this study, but the present and previous [15-19] results on the bcc-rhombohedral phase boundary, the results from previous DAC [1] and SW [16] experiments, the experimental Hugoniots measured by McQueen [37] and by Foster *et al.* [38], and the theoretical Hugoniot calculated by ourselves. A tentative bcc-rhombohedral phase boundary is drawn (dashed blue line) which qualitatively resembles the phase boundary calculated by Wang *et al.* [19] and is consistent with all the results available in the literature [15-19]. Notice that Wang *et al.* [19] predicted the reentrance of the bcc phase at RT around 280 GPa, which is beyond the pressure limit of our experiments. This prediction should be tested by future experiments.

The first comment we would like to make, based on our results, is that both experimental techniques give melting points that are consistent with each other and agree within error bars with our computational study. The simulations determined the melting temperature at higher pressures than those covered in our experiments, the melting T at 390 GPa being equal to 6870±160 K. As in many other metals [39], the experimental melting curve can be described by a Simon-Glatzel equation [40]: $T_M(K) = 2183 \times \left(1 + \frac{P}{32}\right)^{0.46}$, where P is in GPa. For this melting curve, the initial slope, at P = 0 GPa, 31.4 K/GPa, is consistent with 32.6 K/GPa from isobaric-heating measurements [41]. The extrapolation of this melting curve runs parallel to the DFT melting curve with a maximum T difference of 200 K (comparable with error bars of experiments and calculations). Notice that the melting curve reported here is higher than the one determined from DAC experiments using the laser-speckled method [1]. The differences up to 30 GPa are below 300 K, but they increase with increasing P, reaching 1000 K at 80 GPa. In addition, we have noticed that the previous melting curve [1] agrees within 200 K with the temperatures at which the onset of recrystallization is detected in our experiments. Indeed, the lowest temperature where recrystallization is detected (green circles in Fig. 4) follow a P dependence which is very similar to the melting curve previously reported in [1]. Therefore, it is quite likely that the older DAC melting curve is the recrystallization curve of V rather than its melting curve. This recrystallization is probably related to the microstructure formation reported in Mo under HP laser heating at temperatures below its melting curve [9]. The phenomenon is very rapid and would produce movements that the laser speckle method would signify as melting [1], and consequently lead to an underestimation of the melting T. This observation is consistent with the spalling and fragmentation induced in laser-shocked V, causing damage to the surface that resembled melting [42].

Another important observation is that the melting curve reported here is well below the melting curve reported from SW experiments [16]. At 200 GPa, our calculated melting temperature is 2000 K lower than that reported by Dai *et al.* [16]. In addition, the Hugoniot temperatures reported by these authors are much higher than our theoretical Hugoniot, as can be seen in Fig. 4. Our Hugoniot, however, is in perfect agreement with the experiments reported by McQueen [37] and Foster *et al.* [38] (see Fig. 4). Thus, it is reasonable to assume that the melting temperatures inferred by Dai *et al.* [16] have been overestimated. In contrast, their pressures seem to be correct: from the Hugoniot sound velocity measurements they determine the Hugoniot melting pressure of 225 GPa; this is exactly the pressure at which our theoretical Hugoniot crosses our melting curve. Another observation we can make based on our results is that the melting temperature calculated by Landa *et al.* [15] at 182 GPa (8000 K) is also a severe overestimation. In their experiments, Dai *et al.* [16] have determined the temperatures from a grey-body fit to six radiance measurements at discrete wavelengths. Such a method for temperature determination is not as accurate as the determination from a wavelength continuum, as performed by us here. In addition, their measurements can be affected by changes in the electronic properties of V under extreme compression [43], by reflections at interfaces between the sample and the interferometric window used in the experiments, and by the fact that it is not very precisely known how the properties of the window material (e.g., reflectivity) are affected by shock waves. On the other hand, the overestimation of the melting temperature in the calculations reported by Landa *et al.* [15] can be related to the use of a linear muffin-tin orbital method. Indeed, the use of the (rigid) muffin-tin approach to the calculation of the elastic constants of transition metals [44] leads to the overestimation of both $C_{11}$ and $C_{44}$ for V by a factor of ~ 2, hence the shear modulus, G, is overestimated by a factor of ~ 2 as well. Since the melting temperature is proportional to the shear modulus [45], it is reasonable to assume that the use of this method overestimates the melting temperature by a similar factor, which seems to be the case with Landa's value (~ 8000/5000 = 1.6).

From our experiments we have been able to determine the pressure dependence of the unit-cell volume of V following different isotherms. The results are shown in Fig. 5 together with the 1000 K isotherm reported by Crichton *et al.* [46]. We observed no volume discontinuity at the bcc-to-rhombohedral transition (which is consistent with the fact that the bcc-rhombohedral phase boundary is quasi-horizontal, in view of the Clausius-Clapeyron formula, and assuming that the bcc-to-rhombohedral transition is of the 1$^{st}$ order), and thus decided to include the results on both phases in the analysis of the influence of P and T on the volume. We found that our results at RT can be properly described using a third-order Birch-Murnaghan EOS [47]. The fit is shown in the figure. The unit-cell volume at ambient pressure ($V_0$), bulk modulus ($K_0$), and its pressure derivative ($K_0$') were determined to be 13.91(3) Å$^3$, 152(4) GPa, and 5.4(4), respectively. The corresponding value of the second pressure derivative of the bulk modulus [48] is 0.0477(3) GPa$^{-1}$. These values agree within error bars with those reported by Crichton *et al.* [46] and with most of the bulk moduli reported in the literature, which range from 150 to 160 GPa [46]. Only Jenei *et al.* [18] and Ding *et al.* [17] have reported larger values for the bulk modulus (179 and 195 GPa) in highly non-hydrostatic experiments, which suggests that highly non-hydrostatic conditions can considerably affect the compressibility of V, as observed in other materials [49].

Regarding the HT results, Crichton *et al.* [46] have reported a thermal EOS that reproduces accurately their experimental results up to 1000 K and 11.5 GPa. They have used a Berman model [50] in which the dependence of the linear temperature on the bulk modulus and thermal expansion coefficient are assumed, with $\alpha_0 = 4.8(6)\ 10^{-5}\ K^{-1}$, $\alpha_1 = -2.4(9)\ 10^{-8}\ K^{-2}$, and $dK/dT = -0.0446(7)$ GPa $K^{-1}$ [46]. However, these parameters lead to unphysical results if the thermal EOS is extrapolated to the P-T range of our study. For instance, for temperatures higher than 2000 K, the thermal expansion becomes negative, and at pressures higher than 30 GPa the HT isotherms cross the RT isotherm. We have therefore fitted the parameters again, by including all the results shown in Fig. 5. We obtained $\alpha_0 = 4.6(6)\ 10^{-5}\ K^{-1}$, $\alpha_1 = -1.2(5)\ 10^{-8}\ K^{-2}$, and $dK/dT = -0.011(1)$ GPa $K^{-1}$. The constant term of the thermal expansion ($\alpha_0$) agrees with the value reported by Crichton *et al.* [46], but $\alpha_1$ is reduced to half of their value (however, the upper limit of our error bar overlaps with the lower limit of the error bar of the previous results). Regarding dK/dT, our value is one fourth of the previously reported value, but is similar to the values reported for this parameter in most metals [51, 52].

## 4. Conclusions

We have reported an accurate determination of the melting curve and phase diagram of V up to 120 GPa and 4000 K. From the XRD experiments carried out we obtained melting points and the phase boundary between the bcc phase and the rhombohedral phase of V. These results were confirmed by the measurement of thermal plateaus in additional experiments, and by DFT calculations. Our melting curve is somewhat higher than that from the previous DAC experiment, but considerably lower than that reported in SW experiments. Explanations for these discrepancies have been provided. Finally, a PVT EOS for V is generated, which is valid for the P-T range covered in our experiments. These results will contribute to an improvement in the understanding of the HP-HT behavior of transition metals.

**Acknowledgments**

The authors thank the European Synchrotron Radiation Facility and especially the beamline ID27, for the beam-time allocated. D.E. acknowledges the financial support from the Spanish Ministerio de Ciencia, Innovación y Universidades, the Spanish Research Agency, the Generalitat Valenciana, and the European Fund for Regional Development under Grants No. MAT2016-75586-C4-1-P, PGC2018-097520-A-100, and Prometeo/2018/123 (EFIMAT). D.S.P. acknowledges the Spanish government for a Ramon y Cajal RyC-2014-15643 grant. ©British Crown Owned Copyright 2019/AWE. Published with permission of the Controller of Her Britannic Majesty's Stationary Office. Computer simulations were performed on the LANL clusters Pinto and Badger. HC was partly supported under the auspices of the U.S. Department of Energy by Lawrence Livermore National Laboratory under Contract DE-AC52-07NA27344. We would like to extend our thanks to Professor Malcolm McMahon for useful discussions during the writing of this manuscript.

# References

[1] D. Errandonea, B. Schwager, R. Ditz, C. Gessmann, R. Boehler, and M. Ross, Phys. Rev. B **63**, 132104 (2001).

[2] S. Anzellini1, A. Dewaele, M. Mezouar, P. Loubeyre, and G. Morard, Science 340, **464** (2013).

[3] R. Boehler, D. Santamaría-Pérez, D. Errandonea, and M. Mezouar, J. Phys.: Conf. Ser. **121**, 022018 (2008).

[4] T. Sun, J. P. Brodholt, Y. Li, and L. Vočadlo, Phys. Rev. B **98**, 224301 (2018).

[5] D. Errandonea, M. Somayazulu, D. Hausermann, and H. K. Mao, J. Phys.: Condensed Matter **15**, 7635 (2003).

[6] A. Dewaele, M. Mezouar, N. Guignot, and P. Loubeyre, Phys. Rev. Lett. **104**, 255701 (2010).

[7] L. Burakovsky, S. P. Chen, D. L. Preston, A. B. Belonoshko, A. Rosengren, A. S. Mikhaylushkin, S. I. Simak, and J. A. Moriarty, Phys. Rev. Lett. **104**, 255702 (2010).

[8] D. Santamaría-Pérez, M. Ross, D. Errandonea, G. D. Mukherjee, M. Mezouar, and R. Boehler, J Chem Phys. 130, 124509 (2009).

[9] R. Hrubiak, Y. Meng, and G. Shen, Nature Communications **8**, 14562 (2017).

[10] C. Cazorla, M. J. Gillan, S. Taioli, D. Alfè, J. Chem. Phys. **126**, 194502 (2007).

[11] C. J. Wu, P. Söderlind, J. N. Glosli, and J. E. Klepeis, Nature Materials **8**, 223 (2009).

[12] D. Errandonea, Nature Materials **10**, 170 (2009).

[13] M. Ross, D. Errandonea, and R. Boehler, Phys. Rev. B **76**, 184118 (2007).

[14] A. Belonoshko, L. Burakovsky, S. P. Chen, B. Johansson, A. S. Mikhaylushkin, D. L. Preston, S. I. Simak, and D. C. Swift, Phys. Rev. Lett. **100**, 135701 (2008).

[15] A. Landa, P. Söderlind, and L. Yang, Phys. Rev. B 89, 020101(R) (2014).

[16] C. Dai, X. G. Jin, X. M. Zhou, J. Liu, and J. Hu, J. Phys. D: Appl. Phys. **34**, 3064 (2001).

[17] Y. Ding, R. Ahuja, J. Shu, J. P. Chow, W. Luo, and H. K. Mao, Phys. Rev. Lett. **98**, 085502 (2007).

[18] Z. Jenei, H. P. Liermann, H. Cynn, J. H. P. Klepeis, B. J. Baer, and W. J. Evans, Phys Rev B. **83**, 054101 (2011).

[19] Y. X. Wang, Q. Wu, X. R. Chen, H. Y. Geng, Scientific Reports **6**, 32419 (2016).

[20] P. I. Dorogokupets and A. Dewaele, High Press. Res. **27**, 431 (2007).

[21] K. D. Litasov, P. N. Gavryushkin1, P. I. Dorogokupets, I. S. Sharygin1, A. Shatskiy, Y. Fei, S. V. Rashchenko, Y. V. Seryotkin, Y. Higo, K. Funakoshi, and E. Ohtani, J. Appl. Phys. **113**, 133505 (2013).

[22] E. Schultz, M. Mezouar, W. Crichton, S. Bauchau, G. Blattmann, D. Andrault, G. Fiquet, R. Boehler, N. Rambert, B. Sitaud, and P. Loubeyre, High. Press. Res. **25**, 71 (2005).

[23] G. Shen, M. I. Rivers, Y. B. Wang, and S. R. Sutton, Rev. Sci. Instrum. **72**, 1273 (2001).

[24] W. Kraus and G. Nolze, J. Appl. Cryst. **29**, 301 (1996).

[25] D. Errandonea, Phys. Rev. B **87**, 054108 (2013).

[26] A. Dewaele, M. Torrent, P. Loubeyre. and M. Mezouar, Phys. Rev. B **78**, 104102 (2008).

[27] C. K. Singh, B. K. Pande, and A. K. Pandey, AIP Conference Proceedings **1953**, 130002 (2018).

[28] L. Burakovsky, N. Burakovsky, M. J. Cawkwell, D. L. Preston, D. Errandonea, and S. I. Simak Phys. Rev. B **94**, 094112 (2016).

[29] D. Errandonea, J. Phys. Chem. Sol. **70**, 1117 (2009).

[30] V. B. Prakapenka, G. Shen, and L. S. Dubrovinsky, High Temp. High Press. **35**, 2 (2003).

[31] D. Antonangeli, D. L. Farber, A. Bosak, C. M. Aracne, D. G. Ruddle, and M. Krisch, Scientific Reports **6**, 31887 (2016).

[32] P. D. Bogdanoff, B. Fultz, J. L. Robertson, and L. Crow, Phys. Rev. B **65**, 014303 (2001).

[33] R. Boehler, M. Ross, and D. B. Boercker, Phys. Rev. Lett. **78**, 4589 (1997).

[34] O. T. Lord, I. G. Wood, D. P. Dobson, L. Vočadlo, W. Wang, A. R. Thomson, E. Wann, G. Morard, M. Mezouar, and M. J. Walter, Earth Planet. Sci. Lett. **408**, 226 (2014).

[35] A. Dewaele, M. Mezouar, N. Guignot, and P. Loubeyre, Phys. Rev. B **76**, 144106 (2007).

[36] S. Anzellini, V. Monteseguro, E. Bandiello, A. Dewaele, L. Burakovsky, and D. Errandonea, Scientific Report **9** (2019), *in press*.

[37] R. G. McQueen and S. P. Marsh, J. Appl. Phys. **31**, 1253 (1960).

[38] J. M. Foster, A. J. Comley, G. S. Case, P. Avraam, S. D. Rothman, A. Higginbotham, E. K. R. Floyd, E. T. Gumbrell, J. J. D. Luis, D. McGonegle, N. T. Park, L. J. Peacock, C. P. Poulter, M. J. Suggit, and J. S. Wark, J. Appl. Phys. **122**, 023517 (2017).

[39] D. Errandonea, J. Appl. Phys. **108**, 033517 (2010).

[40] F. E. Simon and G. Glatzel, Z. Anorg. Chem. **178**, 309 (1929).

[41] G. R. Gathers, J. W. Shaner, R. S. Hixson and D. A. Young, High T - High P **11**, 653 (1979).

[42] H. Jarmakani, B. Maddox, C. T. Wei, D. Kalantar, and M. A. Meyers, Acta Mater. **58**, 4604 (2010).

[43] S. I. Ashitkov, P. S. Komarov, E. V. Struleva, M. B. Agranat, and G. I. Kanel, JETP Letters **101**, 276 (2015).

[44] Y. Ohta and M. Shimizu, Physica B **154**, 113 (1988).

[45] L. Burakovsky, D. L. Preston and R. R. Silbar, J. Appl. Phys. **88**, 6294 (2000).

[46] W. A. Crichton, J. Guignard, E. Bailey, D. P. Dobson, S. A. Hunt, and A. R. Thomson, High Press. Res. **36**, 16 (2016).

[47 F. Birch, J. Geophys. Res. **83**, 1257 (1978).

[48] D. Errandonea, Ch. Ferrer-Roca, D. Martinez-Garcia, A. Segura, O. Gomis, A. Munoz, P. Rodriguez-Hernandez, J. Lopez-Solano, S. Alconchel, F. Sapina, Phys. Rev. B **82**, 174105 (2010).

[49] D. Errandonea, A. Muñoz, and J. Gonzalez-Platas, J. Appl. Phys. **115**, 216101 (2014).

[50] R. J. Angel, J. Gonzalez-Platas, and A. Alvaro, Z. Kristallogr. **229**, 405 (2014).

[51] S. Anzellini, D. Errandonea, S. G. MacLeod, P. Botella, D. Daisenberger, J. M. De'Ath, J. Gonzalez-Platas, J. Ibáñez, M. I. McMahon, K. A. Munro, C. Popescu, J. Ruiz-Fuertes, and C. W. Wilson, Phys. Rev. Mat. **2**, 083608 (2018).

[52] D. Errandonea, S. G. MacLeod, J. Ruiz-Fuertes, L. Burakovsky, M. I. McMahon, C. W. Wilson, J. Ibañez, D. Daisenberger, and C. Popescu, J. Phys.: Condens. Matter **30**, 295402 (2018).

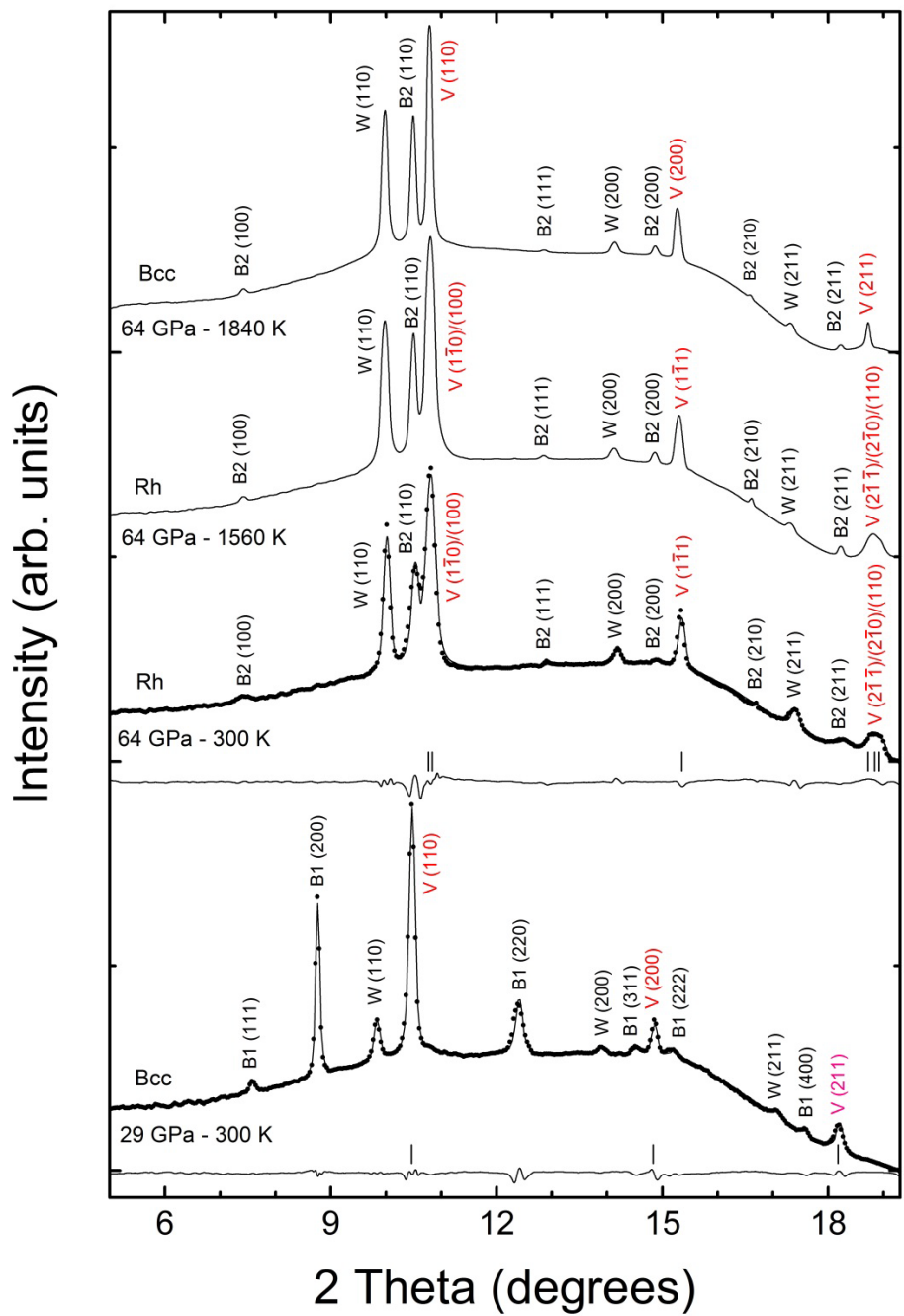

**Figure 1:** (Color online) XRD patterns at selected pressures and temperatures (indicated in the figure). In the two lowest traces, experiments are shown with symbols and Rietveld refinements and residuals are shown with solid lines. The ticks correspond to positions of V peaks. The splitting of V peaks due to the rhombohedral distortion can be clearly seen. All peaks are labeled (V peaks in a different color to facilitate the identification). W identifies peaks from tungsten and B1 and B2 peaks from the different phases of NaCl. Bcc and Rh are used to identify the bcc and rhombohedral phases.

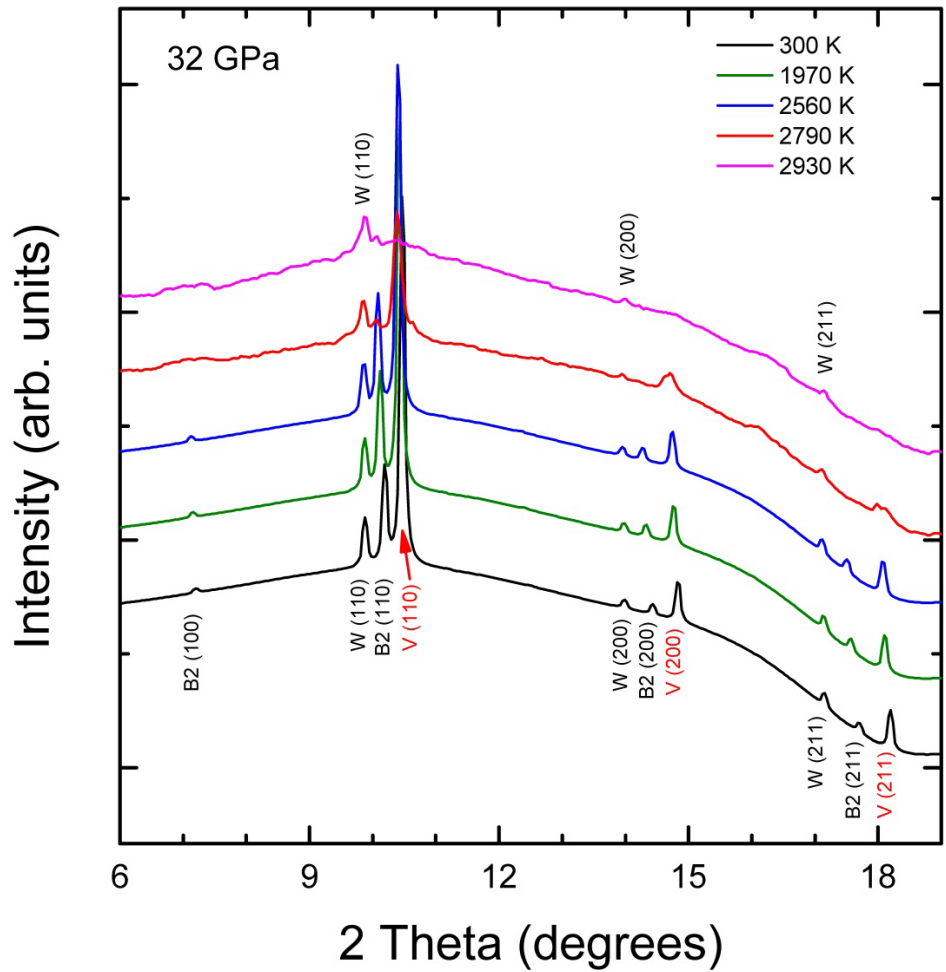

**Figure 2:** (Color online) XRD patterns for a heating run at 32 GPa. The peaks of V, W, and B2 NaCl are identified. Temperatures are given in the figure. The pattern measured at 2790 K corresponds to a temperature where permanent recrystallization of V is observed.

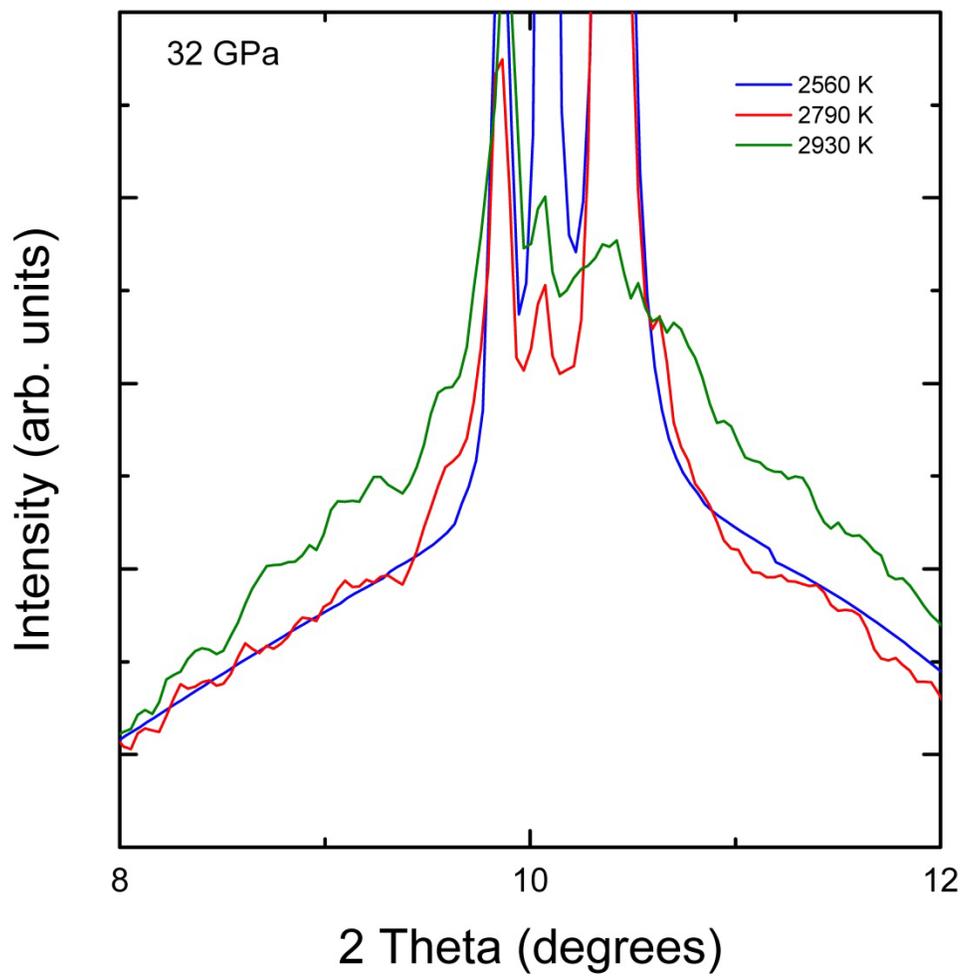

**Figure 3:** (Color online) Zooming in on the region of the XRD patterns measured at 32 GPa to illustrate the disappearance of the (110) peak of V and the increase of the background. Temperatures are indicated in the figure.

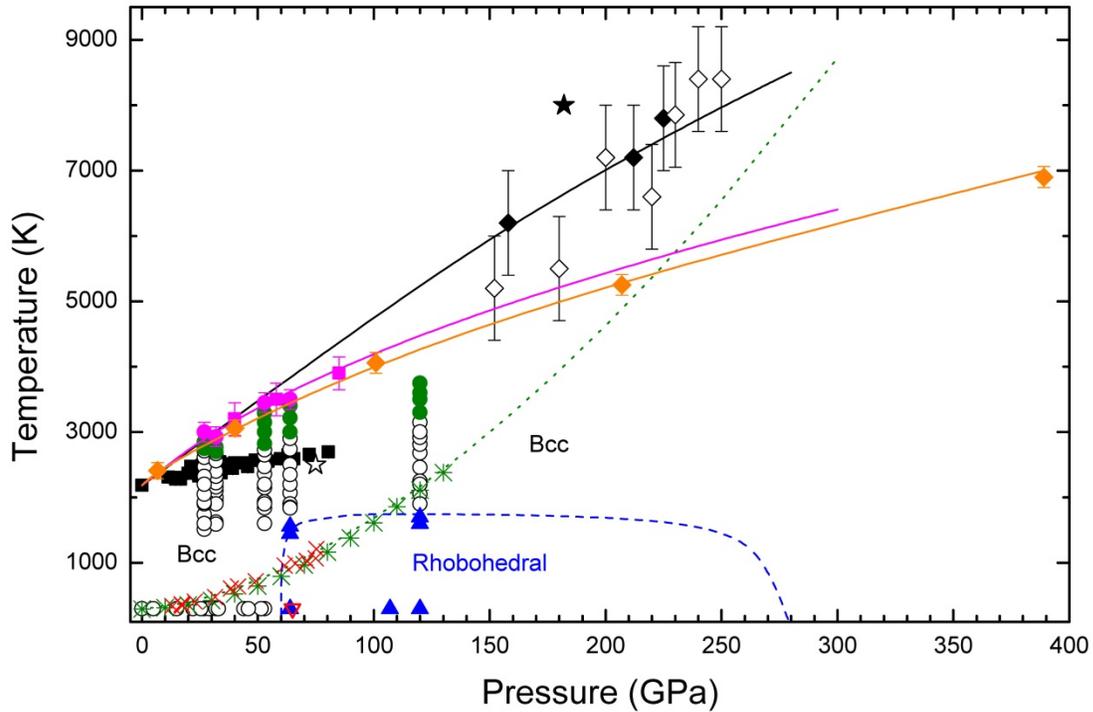

**Figure 4:** (Color online) P-T phase diagram of V. White circles: bcc V. Blue triangles: Rhombohedral V. Upside-down pink triangle: Rhombohedral V from Ref. 17. Blue dashed line is the tentative bcc-rhombohedral phase boundary based upon present and previous studies [17, 19]. White star represents the bcc stability point calculated by Landa *et al.* [15]. Green circles are the points where rapid recrystallization is observed. Pink circles are the melting points determined from present synchrotron experiments. Pink squares are the melting points determined from temperature plateaus. The pink line is the Simon-Glatzel fit to the present melting results. Orange line and diamonds is the calculated melting curve using the Z method. Green asterisks belong to the Hugoniot measured by McQueen [37] and red crosses to the Hugonot measured by Foster *et al.* [38]. Green dotted line is our calculated Hugoniot. White diamonds are the Hugoniot points measured by Dai *et al.* [16]. Black diamonds are the melting points determined by the same authors in SW experiments, and the black line is the melting curve proposed by them. Black squares are the DAC melting results from Mainz [1]. Black star is the melting point calculated by Landa *et al.* [15].

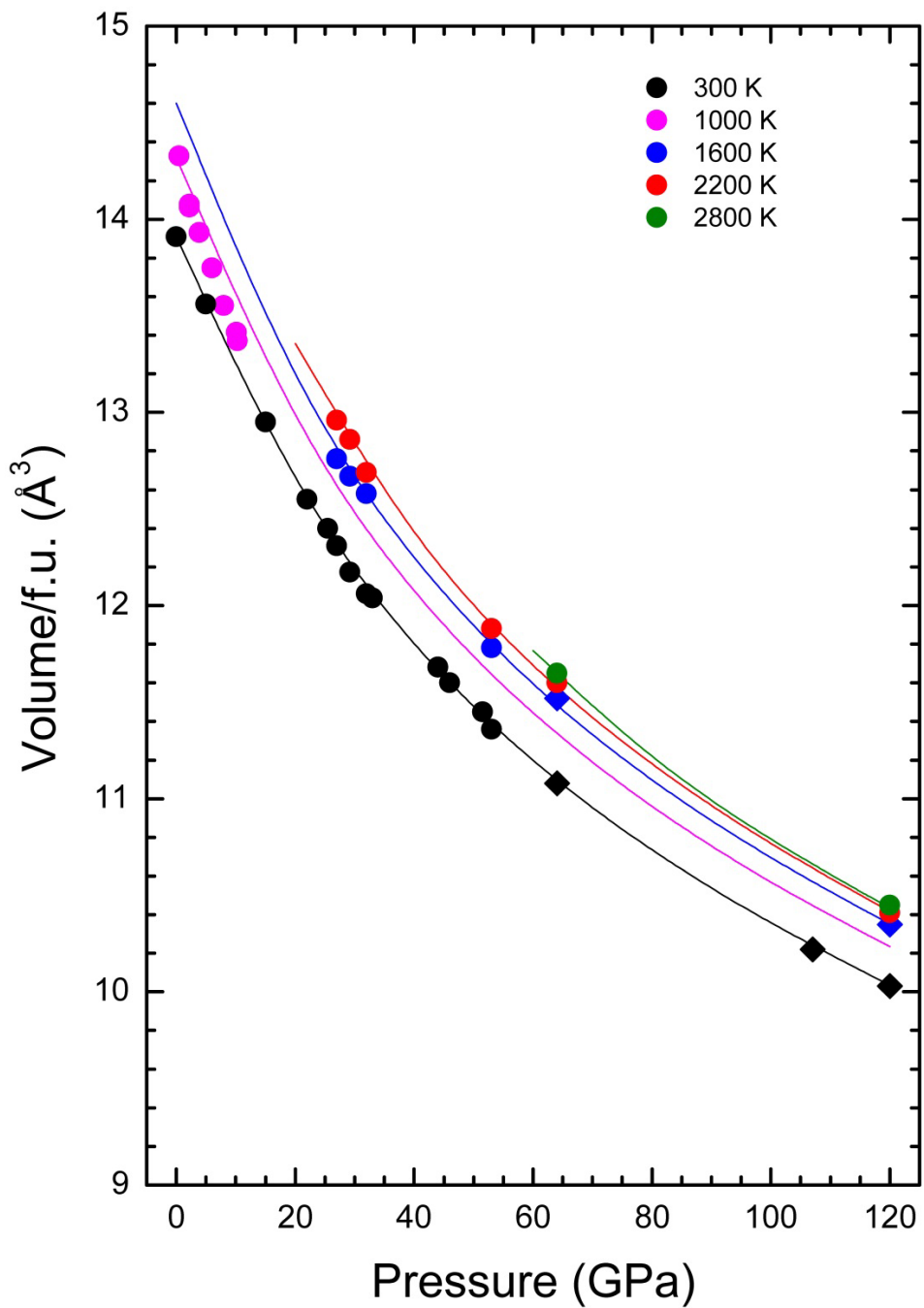

**Figure 5:** (Color online) Pressure evolution of the unit-cell volume per formula unit (f.u.) for different isotherms identified by colors (temperatures are indicated in the figure). Circles correspond to the bcc phase and diamonds to the rhombohedral phase. The results at 1000 K were taken from Ref. 43. The solid lines correspond to the P-V-T EOS reported here.